\documentstyle[12pt,psfig]{article}
\begin{document}

\newcommand{\pronabla}{\stackrel{\rightarrow}{\nabla}}
\newcommand{\conabla}{\stackrel{\leftarrow}{\nabla}}
\newcommand{\bothnabla}{(\pronabla-\conabla)} 
\newcommand{\gaprox}{$ {\raisebox{-.6ex}{{$\stackrel{\textstyle >}{\sim}$}}} $}

\thispagestyle{empty}

\hfill{KRL MAP-230}

\hfill{NT@UW-98-01}

\vfill
\vspace*{24pt}
\begin{center}
{\large\bf Effective Field Theory of Short--Range Forces}

\vspace*{48pt}

{\bf U. van Kolck} 
\vspace*{12pt}

{\sl Kellogg Radiation Laboratory, 106-38}\\
{\sl California Institute of Technology}\\
{\sl Pasadena, CA 91125}\\

and 

{\sl Department of Physics}\\
{\sl University of Washington}\\
{\sl Seattle, WA 98195-1560}

\vspace*{9pt}
{\tt vankolck@krl.caltech.edu}

\vspace*{36pt}

\begin{abstract}
The method of effective field theories (EFTs) is developed for
the scattering of two particles 
at wavelengths which are large compared to the range of their interaction.
It is shown that the 
renormalized EFT is equivalent to the effective range expansion, 
to a Schr\"odinger equation with a pseudo-potential, and to an energy 
expansion of a generic boundary condition at the origin.
The roles of regulators and potentials are also discussed. 
These ideas are exemplified in a toy model.
\end{abstract}

\vspace*{12pt}

\end{center}

\vfill
\newpage
\setcounter{page}{1}
\setcounter{section}{0}
\section{Introduction}

Separation of scales is essential in most problems of physics because it 
allows the selection of relevant degrees of freedom with a perhaps limited 
set of dominant interactions. Quantitative understanding then eventually 
requires
a systematic treatment of the less relevant interactions, and methods to
accomplish this are several. In nuclear physics this problem was faced
early on when it was realized that the deuteron is large compared to the
range of the nuclear force, and also in the analysis of
slow-neutron scattering from bound
protons, where the effects of the nuclear force had to be 
separated from those of the longer-range,
electromagnetic interactions. This interest led to the treatment of 
interactions of short range by the equivalent 
techniques of boundary conditions \cite{Peierls,Breit}, 
pseudo--potentials \cite{Fermi,Breit,Huang}, 
and the effective range expansion \cite{Bethe}.
With time, these techniques came to be replaced by parametrizations of
the nuclear force in terms of (single, sometimes double) meson exchange,
which were perceived as more fundamental. 
Meanwhile, the method of effective field theories (EFTs) has been developed
in particle physics. The idea is to
start from the most general Lagrangian involving the relevant
low-energy degrees of freedom and some chosen symmetries,
short-range dynamics being described by local interactions 
with an arbitrary number of derivatives. Under the assumption that these
local terms have natural size, a power counting argument shows that there
exist an expansion in energy that allows the computation of low-energy 
observables in terms of parameters that summarize the effect of 
short-range interactions. This approach is now viewed as the 
paradigm for understanding the successes of the electroweak theory 
and of chiral Lagrangians.
It is natural to ask what is the connection among these different methods.

In the last few years, the 
application of EFT to the nuclear force problem 
\cite{Weinberg2}
has generated a lot of 
interest (see Ref. \cite{vKolckn} for a review and further references).
To a large extent, the potential usefulness of the approach lies
in the fact that the general chiral 
Lagrangian allows a clear separation between pion effects constrained by
the approximate chiral symmetry of QCD and shorter range effects that 
cannot at present be calculated from the QCD dynamics. Although a 
quantitative description of all 
nucleon-nucleon ($NN$) channels can be accomplished 
up to lab energies of about 100 MeV \cite{Ordonez}
and a qualitative understanding of 
otherwise mysterious nuclear regularities is possible
\cite{vKolck,isoKolck}, 
several questions regarding the scope of such approach
remain unanswered, such as the role of cut-off {\it vs.} dimensional 
regularization, the usefulness of a dibaryon field, and the relation to the 
effective range expansion 
\cite{Kaplan2,david1,cohen,luke2,Lepage,Birse1,Park}.
These issues do not typically arise in particle physics where most 
applications of EFT are to systems where the low-energy domain is perturbative
---that is, involves a finite number of diagrams in the effective theory.
Most of these issues are in fact more general than the $NN$ problem itself, 
being independent of the complications generated by spin-dependent 
interactions, by several partial waves, or by 
(long-range) pion exchange. 
They arise from the necessity of summing up an infinite 
number of diagrams due to infrared quasi-divergences in any EFT
that contains heavy stable particles. 

Here I will therefore restrict myself to a system of 
two identical, stable spinless particles 
at energies much smaller than their mass, their 
internal excitation energy, and
the range of their interaction.
Generalization to spin and non-identical particles is straightforward, and
is  directly relevant to strong interactions
in any nuclear reaction at sufficiently low energies.
In the case of
$NN$ scattering, relevance is limited to center-of-mass momenta much 
smaller than the pion mass.
Generalization to a potential with both long- and short-range components
---appropriate for example to charged systems and to $NN$ strong interactions
at
center-of-mass momenta comparable to the pion
mass--- will be presented in a separate publication.

I will show in detail how the infinite number of interactions in
the EFT can be organized systematically by means of power counting,
in any regularization scheme.
The power counting itself depends on whether the underlying short-range
dynamics is generic, or fine-tuned to display a shallow bound state, or
fine-tuned to display a shallow amplitude zero.
These three cases are treated in Sect. 3, after some generalities
are presented in Sect. 2.
We are going to see that the EFT method, when carried out to the end, is
equivalent to the (generalized) effective range expansion. 
The 
Schr\"odinger equation that results from renormalization 
of the EFT problem will be derived in Sect. 4.
I will show that it involves a delta-function
pseudo-potential of the type 
previously considered in nuclear physics, or, equivalently, boundary 
conditions at the origin. 
I will discuss the relation to the traditional potential approach
in Sect. 5. Necessary restrictions on the regularization scheme
will be presented for the traditional approach to hold. 
A toy potential is played with in Sect. 6 to illustrate some
of the consequences of power counting, before conclusion
in Sect. 7.

Some of these results have been first presented briefly in Ref. 
\cite{vKolckn},
and applied to the three-nucleon problem
in collaboration with P.F. Bedaque and H.-W. Hammer \cite{Bedaque},
with considerable success.
In the case of a shallow bound state, the power counting has subsequently 
been re-discovered within specific regularization schemes
in Refs. \cite{Kaplan4,Gegelia1}, and enlarged to include low-momentum
pions in Ref. \cite{Kaplan4}, while
the equivalence to the effective range expansion was
recently confirmed using renormalization group methods \cite{Birse2}. 
Some of the 
implications of the power counting to the potential approach 
have been checked numerically
for a toy underlying theory in Ref. \cite{steele}.

\section{Generalities}

Consider the case of particles (described by a field $\psi$)
with 3-momenta $Q$ much smaller
than their mass $m$, the mass difference $\Delta$ to their first excited 
state, and the mass $\mu$ of the lightest particle that 
can be exchanged among them. 
The EFT appropriate to these small momenta 
contains only the field $\psi$ as a propagating,
non-relativistic
degree of freedom. The effect of all other states is to generate 
structure and interactions of range 
$\sim 1/\Delta$ or $\sim 1/\mu$,
while relativistic corrections introduce further $\sim 1/m$ terms.
Low-energy $S$-matrix  elements can be 
obtained from an effective Lagrangian involving arbitrarily complicated 
operators of only $\psi$ and its derivatives. Naively, we expect these
derivatives to be associated with factors of $1/m$, $1/\Delta$, or $1/\mu$
and, therefore, 
that the effective Lagrangian can be written as
an expansion in $\partial/(m, \Delta, \mu)$.
More generally, letting $M$ characterize the typical scale of all 
higher-energy effects, we seek an expansion of observables in powers
of $Q/M$.

Because I am restricting myself to soft collisions, the 
relevant symmetry of the EFT 
is invariance under Lorentz transformations of small 
velocity, sometimes referred to as reparametrization invariance 
\cite{luke1}. 
For simplicity, I only consider parity and time-reversal invariant
theories.
The particles
$\psi$ are necessarily non-relativistic, and evolve only forward in time,
in first approximation as static objects. Particle number is also conserved.
By a convenient choice of fields, the effective Lagrangian can be written 
as
\begin{eqnarray}
\cal L & = & \psi^\dagger \left(i\partial_{0}
              +\frac{1}{2m} \vec{\nabla}^{2}
              +\frac{1}{8m^3}\vec{\nabla}^{4}+\ldots\right) \psi
                                                                  \nonumber \\
 &  & - \frac{1}{2}C_0 \psi^\dagger \psi\: \psi^\dagger \psi       \nonumber \\
 &  & - \frac{1}{8} (C_2+C'_2) [\psi^\dagger \bothnabla\psi \cdot
                        \psi^\dagger \bothnabla\psi
                        - \psi^\dagger \psi \:
                          \psi^\dagger \bothnabla^2\psi] \nonumber \\
 &  & + \frac{1}{4} (C_2-C'_2)
                         \psi^\dagger\psi\:\vec{\nabla}^2(\psi^\dagger \psi) 
       +\ldots,                                   \label{lag}
\end{eqnarray}
\noindent
where the $C_{2n}$'s 
are parameters that depend on the details 
of the dynamics of range
$\sim 1/M$. I will restrict myself here to four space-time dimensions,
although extension to three dimensions ---which also has some 
phenomenological interest--- is straightforward. 
In  four space-time dimensions,
$C_{2n}$ has mass dimension $-2(1+n)$.
The ``$\ldots$'' include operators with more derivatives and/or
more fields. 
The latter will not contribute to $\psi\psi$ scattering.

Canonical quantization leads to 
familiar Feynman 
rules. For example, the $\psi$ propagator at four-momentum $p$ is given by
\begin{equation}
S(p^0, \vec{p}) = 
  \frac{i}{p^0-\frac{\vec{p}^{\, 2}}{2m}
              +\frac{\vec{p}^{\, 4}}{8m^3}+\ldots+i\epsilon}
                                        \label{fprop}
\end{equation}
\noindent
and the four-$\psi$ contact interaction by $-i v(p,p')$, with
\begin{equation}
v(p,p')= C_0+ C_2 
(\vec{p}^{\: 2}+\vec{p}\:'^{\, 2})
+ 2C'_2 \vec{p}\cdot\vec{p}\:'+ \ldots, \label{ver}
\end{equation}
\noindent
$\vec{p}$ ($\vec{p}\:'$) being the relative momentum of
the incoming (outgoing) particles. 
If Fourier-transformed to coordinate space, 
these bare interactions consist of delta-functions and 
their derivatives. 

I will concentrate here on the two-particle system 
at energy $E= k^2/m- k^4/4m^3+\ldots$ 
in the center-of-mass frame. 
Conservation of particle number
reduces the $T$-matrix to a sum of bubble diagrams. 
These bubbles will give contributions containing
\begin{equation}
\int \frac{dl^0}{2 \pi} \: S\left( \frac{E}{2} +l^0, \vec{p} + \vec{l}\,\right)
                      S\left( \frac{E}{2} -l^0, -(\vec{p} + \vec{l}\:) \right).
\end{equation}
\noindent
Upon integration over the
zeroth component of the loop momentum, the two particles
evolve formally according to the familiar non-relativistic
Schr\"odinger propagator
\begin{equation}
G_0(l;k)= -\frac{m}{l^2 -k^2 -i\epsilon}         \label{sprop}
\end{equation}
\noindent
in momentum space, or 
\begin{equation}
G_0(r;k)= -imk \frac{e^{ikr}}{4\pi r}           \label{spropcoord}
\end{equation}
\noindent
in coordinate space.
Relativistic corrections can be accounted for by a two-legged vertex
\begin{equation}
u(l;k) = -\frac{l^4-k^4}{8m^3}+ \dots
\end{equation}

Loops are then associated with non-zero integrals of the form
\begin{equation}
I_{2n}(k) = \int \frac{d^3l}{(2 \pi)^3} \: l^{2n} G_0(l;k)
       = -m \int \frac{d^3l}{(2 \pi)^3} \: \frac{l^{2n}}{l^2-k^2 -i\epsilon},
                                                    \label{bubbleany}
\end{equation}
\noindent
with $n\geq 0$ an integer that depends on the number of derivatives
at the vertices.
Such integrals are ultraviolet divergent, which is the field-theoretical
translation of the well-known fact that 
our bare interactions are too singular
for propagation with the Schr\"odinger propagator.
The problem therefore requires regularization and renormalization. 
This introduces a 
mass scale 
$\Lambda$, which might be a cut-off in momentum space or 
a mass scale generated by extending the dimension of space in 
Eq. (\ref{bubbleany}) to a sufficiently small $D-1$ such that the
integral converges ---I will refer to this as a cut-off scale,
although I am {\it not} assuming a particular regularization scheme.
In general,
power-like dependence on $\Lambda$ appears and can be encoded in
\begin{equation}
L_{n} \equiv \int dl \: l^{n-1} \equiv \theta_n \Lambda^n,
\label{divs}
\end{equation}
\noindent
where $\theta_n$ is a number that
depends on the regularization scheme chosen. In 
dimensional regularization with minimal subtraction, for example, 
$\theta_n=0$, but in general $\theta_n$ can be non-vanishing with
either sign. 
The loop integrals are then 
\begin{equation}
I_{2n}(k)  =  -\frac{m}{2\pi^2}\left[\sum_{i=0}^{n} k^{2i} L_{2(n-i)+1} 
                                + i\frac{\pi}{2}k^{2n+1}
      +\frac{k^{2(n+1)}}{\Lambda} R(k^2/\Lambda^2)\right], \label{regbubbleany}
\end{equation}
\noindent
where $R(x)$ is a regularization-scheme-dependent
function that approaches a finite limit
as $x\rightarrow 0$.

The severe $\Lambda$-dependence comes from the region of high momenta
that cannot be described correctly by the EFT; 
it can be removed
by lumping these terms together with the unknown bare parameters
into ``renormalized'' coefficients $C_{2n}^{(R)}$. 
Only the latter are observable.
The regulator-independent, non-analytic piece in
Eq. (\ref{regbubbleany}), on the other hand, is characteristic of loops:
it cannot be mocked up by re-shuffling contact interactions.
From Eq. (\ref{regbubbleany}) we see that we can estimate its effects 
by associating a factor
$m Q/ 4 \pi$ to each loop and a factor $Q$ to each derivative at
the vertices.

\section{The amplitude}

The goal is to find a power counting scheme that justifies an order by 
order truncation of 
the effective theory to a finite number of interactions. 
The observable results of the EFT should be independent of the choice of
$\Lambda$, {\it but only to the accuracy implied by this
power counting scheme.}

All $\psi\psi$ observables can be obtained from the on-shell $T$-matrix
\linebreak
$T_{os}(k, \hat{p}\,'\cdot\hat{p})$.
(Off-shell behavior cannot be separated from three-body force effects,
and plays no role in the EFT treatment of the two-body system.)
Scattering information is encoded in the phase shifts. 
In the case of two non-relativistic particles of equal mass the relation
between the $l$-wave amplitude
$(T_{os}(k))_l$ and the phase shift $\delta_l(k)$ is \cite{Gold}
\begin{eqnarray}
(T_{os}(k))_l & = & \frac{4\pi}{mk} 
      \left( 1+ \frac{k^2}{2m^2}+O\left(\frac{k^4}{m^4}\right)\right) ^{-1}
            (\cot \delta_l -i)^{-1}   \nonumber \\
& = & \int_{-1}^{1} dx \: P_l(x) T_{os}(k, x).
         \label{Tondel}
\end{eqnarray}
\noindent
Here $(1+ k^2/2m^2+O(k^4/m^4))$ stems from the small 
relativistic corrections in the inverse of the density of states.
At sufficiently small energy it is customary to Taylor-expand
\begin{equation}
k^{2l+1} \cot \delta_l = -\frac{1}{a_l} + \frac{r_l}{2} k^2 
                         - P_l r_l^3 k^4 + \dots,
\label{ere}
\end{equation}
\noindent
or alternatively,
\begin{equation}
k^{-(2l+1)} \tan \delta_l = -a_l \left( 1 + \frac{r_l a_l}{2}  k^2 
          - \left(P_l -\frac{a_l}{4r_l}\right) a_l  r_l^3 k^4 + \dots \right).
\label{invere}
\end{equation}
\noindent
This is known as the effective range expansion,
$a_l$ being the $l$-wave scattering length,
$r_l$ the $l$-wave effective range, 
$P_l$ the $l$-wave shape parameter, {\it etc.} 
A real (virtual) bound state can arise as a pole of 
the amplitude (\ref{Tondel}) at imaginary momentum $k=i\kappa$,
$\kappa\geq 0$ ($\leq 0$). If this pole is sufficiently
shallow, Eq. (\ref{ere}) gives $\kappa$ in terms of the 
effective range parameters.

\begin{figure}[t]
\centerline{\psfig{figure=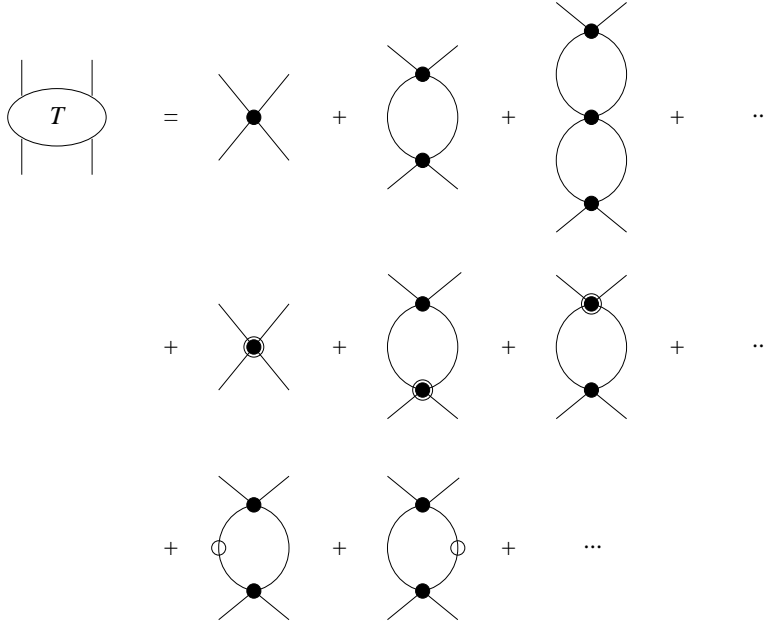,height=3.25in,width=4.00in}}
\caption{First few terms of the 
two-particle amplitude $T$ in a natural EFT.
Two solid lines represent a Schr\"odinger propagator;
a circle on a line represents a $Q^4$ relativistic correction;
a heavy dot stands for a $Q^0$ contact interaction, and
a dot within a circle for a $Q^2$ contact interaction.
\label{fig:natt}}
\end{figure}

The $T$-matrix is given by the diagrams in Fig. \ref{fig:natt}: it is
simply a sum of bubble graphs, 
whose vertices are the four-$\psi$ contact terms that appear 
in the Lagrangian (\ref{lag}). For the first few terms one finds
\begin{eqnarray}
T_{os}(k, \hat{p}\,'\cdot \hat{p}) & = &
     -C_0 [1 +C_0 I_0 + (C_0 I_0)^2+ (C_0 I_0)^3 
                                   + 2C_2 I_2 + \ldots] \nonumber \\
  & & -2C_2 k^2 [1 +C_0 I_0 +\ldots] 
     -C_0^2 [-\frac{L_3}{8\pi^2 m} + \frac{k^2}{2 m^2} I_0] \nonumber \\
  & & -2C'_2k^2 \hat{p}\,'\cdot \hat{p} [1 +\ldots] 
     +\ldots                                 \nonumber \\
& = & -C_0^{(R)} \left\{ 1- \left(\frac{mC_0^{(R)}}{4\pi}ik\right)
    +\left(\frac{mC_0^{(R)}}{4\pi}ik\right)^2
    -\left(\frac{mC_0^{(R)}}{4\pi}ik\right)^3  \right. \nonumber \\
  & & 
\ \ \ \ \ \ \ \ \ \ \ + 2\frac{C_2^{(R)}}{C_0^{(R)}} k^2
 \left[1- 2 \left(\frac{mC_0^{(R)}}{4\pi}ik\right) 
\right] 
+2\frac{C_2^{'(R)}}{C_0^{(R)}} k^2 \hat{p}\,'\cdot \hat{p} 
\nonumber \\
  & & \left. \ \ \ \ \ \ \ \ \ \ \ 
  +\left(\frac{mC_0^{(R)}}{4\pi}ik\right)\frac{k^2}{2m^2} 
+ \ldots \right\}    
                                                        \label{Tpert}
\end{eqnarray}
\noindent
Here dependence on the cut-off was eliminated by 
defining renormalized parameters 
$C_0^{(R)}$, $C_2^{(R)}$, and $C_2^{'(R)}$ 
from the bare parameters 
$C_0(\Lambda)$, $C_2(\Lambda)$, and $C'_2(\Lambda)$:
\begin{eqnarray}
C_0^{(R)} & \equiv & C_0\left[ 1- \frac{mC_0 L_1}{2\pi^2}
                     + \left(\frac{mC_0 L_1}{2\pi^2}\right)^2 
                     - \left(\frac{mC_0 L_1}{2\pi^2}\right)^3   
                    \right.            \nonumber \\
  & &  \ \ \ \ \ \ \left.   - \frac{mC_0 L_3}{2\pi^2}
                 \left(\frac{2C_2}{C_0}+\frac{1}{4m^2}\right)
                 +\ldots\right],    \label{C0R}
\end{eqnarray}
\noindent
or 
\begin{equation}
\frac{1}{C_0^{(R)}} = \frac{1}{C_0} 
+\frac{m}{2\pi^2}\left[ L_1 +\left(\frac{2C_2}{C_0} +\frac{1}{4m^2}\right) 
L_3\right]
    +\ldots,                 \label{1/C0R}
\end{equation}
plus
\begin{equation}
\frac{C_2^{(R)}}{(C_0^{(R)})^2}
     \equiv \frac{C_2}{C_0^2}
      -\frac{m}{4\pi^2} \left(\frac{R(0)}{\Lambda} +\frac{L_1}{2m^2}\right)
            + \ldots,           \label{C2R}
\end{equation}
\noindent 
and 
\begin{equation}
C_2^{'(R)}\equiv C_2^{'} +\ldots  \label{C2'R}
\end{equation}
\noindent
Note that I chose to absorb a finite piece 
$-m (C_0^{(R)})^2 R(0)/4\pi^2 \Lambda$ 
in $C_2^{(R)}$. Other terms coming from $R(k^2/\Lambda^2)$ are 
$\propto k^4$ or higher powers, and cannot be separated from higher-order
contact interactions that I have not written down explicitly.

It is important to realize that Eq. (\ref{Tpert}) consists of 
{\it two} different expansions: a loop expansion and an expansion 
in the number of insertions of derivatives at the vertices or particle 
lines. The derivative expansion depends on the relative size
of the coefficients $C_{2n}^{(R)}$, 
for example $C_2^{(R)}Q^2/C_0^{(R)}$.
The loop expansion is governed by 
$mC_0^{(R)}Q/4\pi$.
The sizes of terms in both expansions thus depend on the 
absolute size of the  $C_{2n}^{(R)}$'s. 
Three different cases will be considered in turn.

\subsection{Natural EFT}

Let us consider first 
the simplest type of underlying theory: one that is 
natural.
This is a theory with a single mass scale $M$ and no fine-tuning.
We expect all parameters to scale with $M$
according to their mass dimension.
It is convenient, however, to factor in a $4\pi/m$: I write
$C_{2n}^{(')(R)}= 4\pi \gamma_{2n}^{(')}/mM^{2n+1}$ with 
$\gamma_{2n}^{(')}$ dimensionless
parameters of $O(1)$.

In this case, the derivative expansion is in 
$C_2^{(R)} Q^2/C_0^{(R)}\sim (Q/M)^2$, while
the loop expansion is in $m Q C_0^{(R)}/4\pi \sim Q/M$.
The $T$-matrix in Eq. (\ref{Tpert}) is thus a simple
expansion in powers of $Q/M$.

We can complete the renormalization procedure by relating the 
parameters of the EFT to observables.
Up to higher-order terms, 
Eq. (\ref{Tpert}) is equivalent to
\begin{equation}
T_{os}(k, \hat{p}\,'\cdot \hat{p}) = (T_{os}^{(0)}(k))_0 
-2C_2^{'(R)}k^2 \hat{p}\,'\cdot \hat{p} + O((4\pi/mM)(Q/M)^4),
\label{renpTon}
\end{equation}
\noindent
where
\begin{equation}
(T_{os}^{(0)}(k))_0 =
- \left[ \frac{1}{C_0^{(R)}} -2 \frac{C_2^{(R)}}{(C_0^{(R)})^2} k^2 
         +\frac{imk}{4\pi} \left(1+\frac{k^2}{2m^2}\right)\right]^{-1}
\left(1+O((Q/M)^4)\right).
                                      \label{renpTonS}
\end{equation}

This is in the form of an effective range expansion in each partial wave.
(Note in particular that the (non-analytic) relativistic corrections
come out in agreement with Eq. (\ref{Tondel}), as they should.)
In the $S$-wave we obtain
a scattering length
\begin{equation}
a_0= \frac{mC_0^{(R)}}{4\pi},                      \label{sscatlen}
\end{equation}
\noindent
and an effective range
\begin{equation}
r_0 = \frac{16\pi}{mC_0^{(R)}} 
      \left(\frac{C_2^{(R)}}{C_0^{(R)}}+\frac{1}{4m^2}\right). 
                                                    \label{seffran}
\end{equation}
\noindent
In the $P$-wave, we find
a scattering volume
\begin{equation}
a_1= \frac{mC_2^{'(R)}}{6\pi}.                      \label{pscatlen}
\end{equation}
\noindent
Higher moments in Eq. (\ref{ere})
can likewise be obtained from
higher-order terms in the $Q/M$ expansion 
of the amplitude. 
For example, the shape parameter is
\begin{equation}
P_0 r_0^3 = \frac{16\pi}{mC_0^{(R)}} 
          \left(\left(\frac{C_2^{(R)}}{C_0^{(R)}}\right)^2
            -\frac{C_4^{(R)}}{C_0^{(R)}}
            +\frac{C_2^{(R)}}{4m^2 C_0^{(R)}}
            +\frac{3}{32m^4}\right),
                                                    \label{sshaparam}
\end{equation}
\noindent
where $C_4$ represents a certain combination of $Q^4$ contact interactions.
We see that an $l$-wave effective range
parameter of mass dimension $\delta$ 
is 
\linebreak 
$O(1/(2l+1)M^{\delta})$. 
This scaling of effective range parameters is indeed what one gets in 
simple potential models, like a square well of  range $R \sim 1/M$ and
depth $V_0 \sim M$, as we will see in Sect. 6.
It is a simple manifestation of the existence of a single mass scale $M$.
The expansion parameter $Q/M$ can then be written alternatively as
$Qa_0 \sim Qr_0 \sim \ldots$
Note, moreover, that unitarity effects in an $l$-wave are further
suppressed by a factor of $2l+1$, being
$\sim Q^{2l+1}a_l\sim Q^{2l+1}/(2l+1)M^{2l+1}$.

The assumption of naturalness thus implies a perturbative amplitude.
Since terms in the $Q/M$ expansion 
of the amplitude are in correspondence 
to terms in the derivative expansion of the effective Lagrangian,
the accuracy of description of low-energy data can be improved systematically
by considering higher-order terms in the Lagrangian (\ref{lag}).
Not surprisingly, the EFT here is quite boring:
it can only describe scattering; 
bound states, if they
exist, have typical momenta $\kappa \sim M$, and are outside the region of 
validity of the expansion (\ref{Tpert}). 

\subsection{An unnatural EFT}

In nuclear or molecular physics, we are also interested in situations 
where there are shallow bound states. 
A (real or virtual) bound state
appears at threshold when a parameter $\alpha$ 
of the underlying theory takes a critical value $\alpha_c$. 
A shallow bound state means that $\alpha$ is close to $\alpha_c$.
That is, the underlying theory is so fine-tuned that it has {\it two}
distinct scales: the obvious $M$ and another 
$\aleph = |\alpha/\alpha_c -1| M \ll M$
induced by the fine-tuning.
I will consider the case of an $S$-wave shallow bound state
by taking  
$C_{2n}^{(R)}= 4\pi \gamma_{2n}/m \aleph (M \aleph)^n$ 
and $C_{2n}^{'(R)}= 4\pi \gamma'_{2n}/m M^{2n+1}$  with 
$\gamma_{2n}^{(')}$ dimensionless
parameters of $O(1)$. 
This of course 
recovers the previous, natural scenario when $\alpha$
is tuned out of $\alpha_c$, and $\aleph$ becomes comparable to $M$.

Now one can see that the terms in $C_0^{(R)}$ in Eq. (\ref{Tpert}) are
the dominant ones at each order in $k$.
For example, the two terms contributing to the coefficient of 
$k^2$, one from the loop expansion, the other from the derivative
expansion, are order $1/\aleph^2$ and $1/\aleph M$, respectively. 
The loop and derivative expansions contained in Eq. (\ref{Tpert}) 
are still in the same combination of parameters as before.
Now, however,
$m Q C_0^{(R)}/4\pi\sim Q/\aleph$, 
while 
$C_2^{(R)} Q^2/C_0^{(R)}\sim Q^2/\aleph M$. 

For $Q\ll \aleph$, the theory is still perturbative, now in $Q/\aleph$.
The situation is similar to a natural theory, except that
perturbation theory breaks down at 
the smaller scale $\aleph$,  much before 
the scale $M$ associated with short-range states.

\begin{figure}[t]
\centerline{\psfig{figure=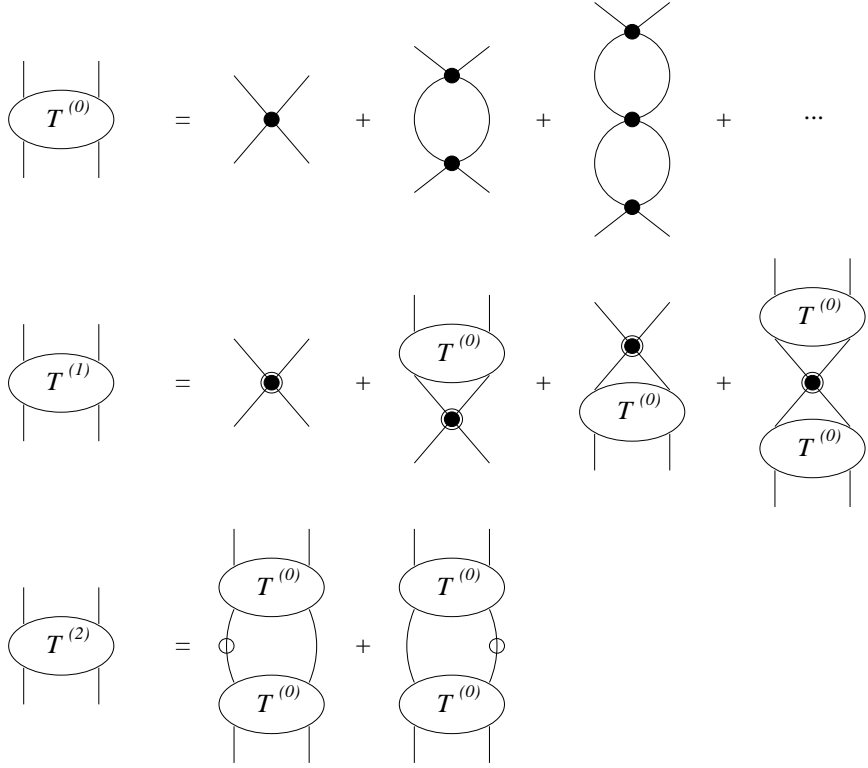,height=4.0in,width=4.50in}}
\caption{First three orders of the
two-particle amplitude $T$ in an EFT with a shallow bound state.
Two solid lines represent a Schr\"odinger propagator;
a circle on a line represents a $Q^4$ relativistic correction;
a heavy dot stands for a $Q^0$ contact interaction, and
a dot within a circle for a $Q^2$ contact interaction.
\label{fig:unnat1t}}
\end{figure}

As $Q$ becomes comparable with $\aleph$,
the most important terms come
from the loop expansion.
The great advantage of short-range interactions is that we can perform
a summation of these terms analytically.
The main contribution to the bubble comes from the 
Schr\"odinger propagator $G_0$ with the $Q^0$ term in the vertex $v(p,p')$.
The bubbles summing to a geometric series, 
one gets as
the leading order amplitude in Fig. \ref{fig:unnat1t}
\begin{equation}
T_{os}^{(0)}(k)  =  -\left( \frac{1}{C_0}- I_0(k)\right)^{-1},  
                                                        \label{Tzero}
\end{equation}
\noindent
which is of $O(4\pi/m (\aleph + Q))$.

Higher-order terms in the derivative expansion can now be accounted for
perturbatively ---see Fig. \ref{fig:unnat1t}.
The first order correction
comes from
one insertion of 
the $C_2$ term in the vertex. We find
\begin{equation}
T_{os}^{(1)}(k)  =  
      -2 \frac{C_2}{C_0^2} 
      \left(k^2 -\frac{m}{2\pi^2}L_3\right) (T_{os}^{(0)}(k))^2
                                                 \label{T1},
\end{equation}
\noindent
which is of $O(Q^2/M(\aleph + Q))$ relative to $T_{os}^{(0)}(k)$ in 
Eq. (\ref{Tzero}).
A similar procedure can be followed in higher orders.
For example, the next corrections come in the $S$-wave.
There are two 
corrections which are 
nominally of 
\linebreak
$O(Q^4/M^2 \aleph (\aleph + Q))$ relative to leading:
{\it(i)} one insertion of a 
combination ---denoted by $C_4$--- of $Q^4$ terms in the Lagrangian
(\ref{lag});
{\it(ii)} two insertions of $C_2$.
There is also one insertion of the relativistic correction in the propagator,
which is of $O(Q^3/m^2(\aleph + Q))$.
The sum of these three corrections I call
$T_{os}^{(2)}(k)$.
The first $P$-wave contribution, one insertion of $C'_2$,
starts at $O(Q^2 (\aleph + Q)/M^3)$:
\begin{equation}
T_{os}^{(3p)}(k, \hat{p}\,'\cdot \hat{p}) = 
                              -2C'_2 k^2 \hat{p}\,'\cdot \hat{p}. 
                                                   \label{T2p}
\end{equation}

The amplitude is
\begin{eqnarray}
T_{os}(k, \hat{p}\,'\cdot \hat{p}) & = & 
       T_{os}^{(0)}(k)+  T_{os}^{(1)}(k) +T_{os}^{(2)}(k) +\dots
       +T_{os}^{(3p)}(k, \hat{p}\,'\cdot \hat{p}) +\ldots \nonumber \\
 & = & (T_{os}^{(0)}(k))_0 
       -2C_2^{'(R)} k^2 \hat{p}\,'\cdot \hat{p}
       +O((4\pi/m\aleph)(\aleph/M)^4);
\end{eqnarray}
\noindent
the $S$-wave component can be rewritten, up to higher order terms, as
\begin{eqnarray}
(T_{os}(k))_0 & = & [ (T_{os}^{(0)}(k))^{-1} 
  +T_{os}^{(1)}(k) (T_{os}^{(0)}(k))^{-2}
   +T_{os}^{(2)}(k) (T_{os}^{(0)}(k))^{-2} +\ldots]^{-1}
        \nonumber \\
 & = & - \left[ \frac{1}{C_0^{(R)}} -2 \frac{C_2^{(R)}}{(C_0^{(R)})^2} k^2 
         +4\left(\frac{(C_2^{(R)})^2}{(C_0^{(R)})^3}
            -\frac{C_4^{(R)}}{(C_0^{(R)})^2}\right)k^4  \right.    \nonumber \\
 & &   \left.  \ \ \ \ \
       +\frac{imk}{4\pi} \left(1+\frac{k^2}{2m^2}\right)\right]^{-1} 
         \left(1+O((\aleph/M)^3)\right).
\label{renTon}
\end{eqnarray}
\noindent 
As before, cut-off dependence of integrals and bare parameters was lumped 
together in renormalized parameters.
Here for simplicity I took $Q\sim \aleph$ in displaying
higher orders. 
This amplitude is nominally correct including
$O((4\pi/m\aleph)(\aleph/M)^2)$ only. I will argue soon that it is
actually correct including $O((4\pi/m\aleph)(\aleph/M)^3)$.

By re-summing the largest terms,
we have produced a new expansion where the leading amplitude is
of $O(4\pi/ m (\aleph+Q))$ and corrections
go as $Q^2/M(\aleph +Q)$ or similar combinations. 
For $Q\ll \aleph$, we regain the perturbative expansion.
The difference is that for $Q\sim \aleph$, 
although all the terms in the loop expansion are of the
same order ($O(4\pi/ m \aleph)$), corrections
from derivatives go in relative powers of
$\aleph/M \ll 1$ and do {\it not} have to be included all at once:
they can be accounted for systematically, at each order in $\aleph/M$. 
As we consider larger $Q$, $Q\gaprox \aleph$, 
the leading terms become 
$O(4\pi/ m Q)$  and corrections get relatively more important,
growing in powers of $Q/M$. The new expansion fails only at momenta 
$\sim M$, as desired.

What has been achieved that is new compared to the natural theory 
is that the low-energy $S$-wave bound state can arise as a pole of 
the amplitude (\ref{renTon}) within the range of validity of the EFT,
at $k=i\kappa$,
\begin{eqnarray}
\kappa & = & \frac{4\pi}{mC_0^{(R)}} 
       \left(1 + 2\frac{C_2^{(R)}}{C_0^{(R)}} \kappa^2 + O(\kappa^3)\right)
                                                  \nonumber \\
       & = & \frac{4\pi}{mC_0^{(R)}} 
       \left(1 + 2\frac{C_2^{(R)}}{C_0^{(R)}}
                  \left(\frac{4\pi}{mC_0^{(R)}}\right)^2 
                                  + O((\aleph/M)^2)\right),
\label{boundmom}
\end{eqnarray}
\noindent
that is, $\kappa\sim \aleph$.
$C_0^{(R)} > 0$ ($<0$) implies $\kappa > 0$ ($<0$) and 
represents a real (virtual) bound state. 
(It can be checked easily that the residue of $i$ times the $S$-matrix
at this pole is indeed positive (negative) if $\kappa > 0$ ($<0$).)
The binding energy is $B= \kappa^2/m + O(\aleph^4/M^3)$, which
to this accuracy is, of course, 
just the usual effective range relation among $B$, $a_0$ and $r_0$.
It is clear that the bound state can be treated in a systematic 
expansion in $\aleph/M$.

Now, Eq. (\ref{renTon}) has the same structure as Eq. (\ref{renpTonS}),
the difference resting on the relative orders of the various
terms. 
There is no change in the formulas for 
$l(>0)$-wave parameters ---such as Eq. (\ref{pscatlen}) for the $P$-wave
scattering volume--- and they scale with $M$ as in the natural scenario.
The $S$-wave is trickier.
The scattering length $a_0$ is given by the same Eq. (\ref{sscatlen})
as before, 
but it now scales with
$\aleph$ rather than $M$, $a_0= O(\aleph^{-1})$.
The formula for the 
effective range $r_0$ is still Eq. (\ref{seffran}), but
the two terms come at different orders. 
The main contribution originates in the contact interactions and 
scales with $M$, so that $r_0 =O(M^{-1})$. The 
relativistic correction, on the other hand, 
is smaller 
by $\sim M\aleph/m^2\sim \aleph/M$; this is 
what justifies the neglect of relativistic corrections
in many situations and the usefulness
of a non-relativistic framework in shallow-bound-state problems.
An underlying theory consisting
of a non-relativistic potential can thus serve as a test
for the scalings of $a_0$ and $r_0$. Indeed, it has been explicitly
shown \cite{lekner} that if the underlying theory consists
of a non-relativistic potential of strength $\alpha$ and range $1/M$, 
then near critical binding effective range parameters behave 
precisely in way derived above. 
A square-well example will be shown in Sect. 6.

As for the shape parameter $P_0$ of Eq. (\ref{sshaparam}),
to the order we have worked so far we recover only the interaction
pieces:
$P_0= ((\gamma_2/\gamma_0)^2-\gamma_4/\gamma_0)M/16\aleph$. 
The problem is, such an $M/\aleph$ behavior is
seen neither in
phenomenological analyses nor in models. This implies that
$\gamma_4$ is $O(1)$ alright, but  
must be precisely $\gamma_4= \gamma_2^2/\gamma_0 + O(\aleph/M)$
so that $P_0$ be $O(1/16)$.
In other words, although each interaction term in Eq. (\ref{sshaparam})
comes at
relative 
$O(Q^4/M^2\aleph(\aleph+Q))$ 
\linebreak
in the amplitude, they are correlated
and their sum is only of relative 
\linebreak
$O(Q^4/M^3(\aleph+Q))$.

We conclude that the shape parameter appears at the same order 
as the first $P$-wave contribution for $Q \gaprox \aleph$. 
Eqs. (\ref{renTon}) and (\ref{boundmom}) are therefore good up to order
$O(Q^2(\aleph+Q)/M^3)$, as advertised. 
For $Q\sim \aleph$, this means that
the first {\it three} orders of the expansion of the amplitude 
---$O(4\pi/m\aleph)$, $O(4\pi/mM)$, and $O(4\pi\aleph/mM^2)$,
depicted in Fig. \ref{fig:unnat1t}---
are pure $S$-wave and given solely by $C_0^{(R)}$ and $C_2^{(R)}$,
which can be determined from $a_0$ and $r_0$.
We can write 
\begin{equation}
T_{os}(k)  =  
    - \left[ \frac{1}{C_0^{(R)}} -2 \frac{C_2^{(R)}}{(C_0^{(R)})^2} k^2 
    +\frac{imk}{4\pi} \left(1+\frac{k^2}{2m^2}\right) \right]^{-1}  
\left(1+ O(\aleph^3/M^3)\right). \label{renTon3}
\end{equation}

It is easy to verify that,  
in order to avoid $M/\aleph$ enhancements in higher effective
range parameters, there must
be such correlations in higher order coefficients as well:
$\gamma_{2n}= \gamma_2^{n}/\gamma_0^{n-1} + O(\aleph/M)$,
or $C_{2n}^{(R)}= C_0^{(R)}(C_2^{(R)}/ C_0^{(R)})^n +$ smaller terms. 
Although not necessary, it is possible to reorganize the EFT in order to 
account explicitly for the dominant correlations.
Because they
form a geometric series in $C_2^{(R)}/C_0^{(R)}$,  
we can sum the dominant correlation terms 
as we have done above with the $C_0^{(R)}$ contributions themselves.
This procedure 
generates a four-$\psi$ interaction of the type
$C_0^{(R)}/(1-2C_2^{(R)}k^2/C_0^{(R)})$;
Eq. (\ref{renTon3}) follows immediately. 

Now, this new re-summation resembles the exchange of an $s$-channel particle, 
$\sigma (g^{(R)})^2/(E - \Delta^{(R)})$ at center-of-mass energy $E=k^2/m$,
provided 
{\it (i)} the sign $\sigma$ of its kinetic term
and its coupling $g^{(R)}$ to $\psi$
satisfy 
\begin{equation}
\sigma (g^{(R)})^2  = - \frac{(C_0^{(R)})^2}{2mC_2^{(R)}},
\end{equation}
\noindent
that is $(g^{(R)})^2/2\pi =O(M/m^2)$; and {\it (ii)} 
its mass be $2m$ plus 
\begin{equation}
\Delta^{(R)}= \frac{C_0^{(R)}}{2mC_2^{(R)}},
\end{equation}
that is, $\Delta^{(R)}=O(M \aleph/2m)$.
This justifies the 
suggestion by Kaplan \cite{david1} 
that one can use a new EFT which involves, besides particles,
also a dibaryon-baryon field with the quantum numbers of the shallow
(real or virtual) bound state.  
In the case under consideration, we are talking a
scalar field, which I denote $T$.
The leading terms in the 
most general Lagrangian consistent with the same symmetries as before
are
\begin{eqnarray}
\cal L & = & \psi^\dagger \left(i\partial_{0}
                           +\frac{1}{2m}\vec{\nabla}^{2}
                          +\frac{1}{8m^3}\vec{\nabla}^{4}+\ldots\right)\psi
                          \nonumber \\
       & & +\sigma  T^\dagger\left(i\partial_{0}
                              +\frac{1}{4m}\vec{\nabla}^{2}
                             -\Delta+\ldots\right)T 
    -\frac{g}{\sqrt{2}} (T^\dagger \psi\psi+\mbox{h.c.})
        +\ldots                       \label{translag}
\end{eqnarray}
\noindent
Note the sign of the kinetic term: the bare $T$ is a ghost
(normal) field if 
\linebreak
$C_2^{(R)} > 0$ ($< 0$), but this is not a problem
because this field does not correspond to an asymptotic state.
``\ldots'' stand for terms with more derivatives,
which are suppressed by powers of $\aleph/M$.
Effects of non-derivative four-$\psi$ contact term
can be absorbed in 
$g^2/\Delta$ and terms with more derivatives.

\begin{figure}[t]
\centerline{\psfig{figure=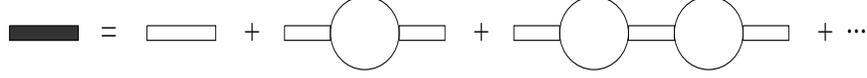,height=0.4in,width=4.50in}}
\caption{The dressed dibaryon propagator:
a bar is the bare dibaryon propagator, 
while the bubbles represent two-particle propagation.
\label{fig:transprop}}
\end{figure}

The coupling to two-particle states dresses the dibaryon propagator.
The dressed propagator contains bubbles as  
in Fig. \ref{fig:transprop}, plus insertions
of relativistic corrections.
This amounts to a self-energy contribution 
proportional to the bubble integral. 
As we know, this integral is ultraviolet 
divergent and requires
renormalization of the parameters of the Lagrangian (\ref{translag}).
Relativistic corrections can be accounted for as in the EFT
without the dibaryon.
The dibaryon propagator in its center-of-mass 
($p^0=k^2/m -k^4/4m^3+ \ldots$)
has the form
\begin{eqnarray}
S(p^0, \vec{0}\,) & = & \sigma \frac{i}{p^0- \Delta
             -\sigma g^2 I_0(\sqrt{mp^0})} \nonumber \\
& = & \sigma\frac{(g^{(R)})^2}{g^2} 
      \frac{i}{\frac{k^2}{m}- \Delta^{(R)}
    + \frac{\sigma m (g^{(R)})^2}{4\pi} ik (1+\frac{k^2}{2m^2}) 
    +\ldots +i\epsilon},
                                   \label{Tprop}
\end{eqnarray}
\noindent
where
\begin{equation}
\frac{\Delta^{(R)}}{(g^{(R)})^2} = \frac{\Delta}{g^2}
              - \frac{m}{2\pi^2} \left( L_1+\frac{L_3}{4m^2} \right) 
                                   +\ldots
\end{equation}
\noindent
and
\begin{equation}
\frac{1}{(g^{(R)})^2} = \frac{1}{g^2}  
 +\frac{m^2}{2\pi^2} \left( \frac{R(0)}{\Lambda}+\frac{L_1}{2m^2}\right)
                        +\ldots
\end{equation}

\begin{figure}[b]
\centerline{\psfig{figure=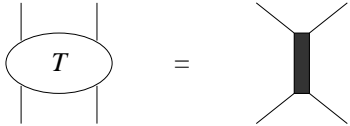,height=0.65in,width=1.8in}}
\caption{The two-particle amplitude $T$ in the EFT with a dibaryon field.
\label{fig:transt}}
\end{figure}

In leading orders, the amplitude in this EFT is ---see Fig. \ref{fig:transt}---
\begin{eqnarray}
T_{os}^{(0)}(k) & = & i g^2 S(p^0, \vec{0}\,) \nonumber \\
& = & - \left[-\frac{\sigma \Delta^{(R)}}{(g^{(R)})^2} 
               + \frac{\sigma }{m(g^{(R)})^2} k^2
    + \frac{imk}{4\pi} \left(1+\frac{k^2}{2m^2}\right) \right]^{-1} 
    \left(1+\ldots\right). \nonumber \\
\label{Ttrans}
\end{eqnarray}

The result is identical to Eq. (\ref{renTon3}), as it should.
This is a more direct way to re-derive the
effective range expansion, but
proceeding this way mixes different orders
in the $\aleph/M$ expansion.
The non-relativistic dibaryon includes the first two orders in 
$\aleph/M$ correctly; the addition of relativistic corrections
regains the first three orders. In either case a number
of higher-order terms is included as well, but this is
irrelevant for the expansion is in a small parameter.  
Corrections stemming from higher-derivative operators
can be accounted for perturbatively, analogously to
what we did in Eq. (\ref{renTon}).
They will contribute to the $\psi \psi$ amplitude starting at
relative $O((\aleph/M)^3)$.

The same real or virtual bound state as before exists near zero energy.
The field $T$ corresponds to a light (ghost or normal) 
state of mass $\Delta^{(R)} \sim \aleph$ relative to $2m$.
It serves to introduce through its bare propagator two
bare poles at the $k=\pm \sqrt{m\Delta}$.
(In dimensional regularization with minimal
subtraction, for example, these bare poles are
on the real (imaginary) axis for $a_0 r_0>0$ ($<0$),
at the geometric mean between the two scales $\aleph$ and $M$.)
This is an undesired pole structure,
but fortunately it changes character as the amplitude gets dressed. 
Upon dressing, the poles move 
to the imaginary axis. One ends up at the position
of the shallow bound state, 
\begin{eqnarray}
\kappa_s = \kappa & = & - \sigma \frac{4\pi\Delta^{(R)}}{m(g^{(R)})^2} 
       \left(1 + \frac{1}{m\Delta^{(R)}} \kappa^2 + O(\kappa^3)\right)
                                                  \nonumber \\
       & = & - \sigma \frac{4\pi\Delta^{(R)}}{m(g^{(R)})^2} 
       \left(1 + \frac{1}{m\Delta^{(R)}}
                  \left(\frac{4\pi\Delta^{(R)}}{m(g^{(R)})^2}\right)^2 
                                  + O((\aleph/M)^2)\right),
\label{shallowboundmom}
\end{eqnarray}
\noindent
that is, $\kappa_s =O(\aleph)$; $\kappa_s>0$ ($<0$)
for $a_0>0$ ($<0$).
The other pole ends up at
\begin{eqnarray}
\kappa_d & = & - \sigma \frac{m^2(g^{(R)})^2}{4\pi} 
       \left(1 +\sigma \frac{4\pi\Delta^{(R)}}{m(g^{(R)})^2} 
                       \frac{1}{\kappa^2} + O(\kappa^{-3})\right)
                                                  \nonumber \\
       & = & - \sigma \frac{m^2(g^{(R)})^2}{4\pi} 
        \left(1 +\sigma  m\Delta^{(R)} 
                   \left(\frac{m^2(g^{(R)})^2}{4\pi}\right)^2
                         + O((\aleph/M)^2)\right),
\label{deepboundmom}
\end{eqnarray}
\noindent
that is, $\kappa_d =O(M)$; $\kappa_d>0$ ($<0$)
for $r_0>0$ ($<0$).
   
Note that the propagator $S$ of Eq. (\ref{Tprop}) is not simply a normalized
particle propagator.
The sign of the residue of $-iS$ at each of the poles depends on
the signs of $a_0$, $r_0$, and $(g^{(R)})^2/g^2$. The latter depends
on the regularization scheme, but this dependence of course 
disappears from the sign of the residue of $i$ times the $S$-matrix.
One finds that the latter is
positive (negative) at $k=i\kappa_s$ for $a_0>0$ ($<0$),
and negative (positive) at $k=i\kappa_d$ for $r_0>0$ ($<0$).
So the shallow pole corresponds to a regular (composite) particle,
while the deep pole corresponds to a ghost. 
In any case, 
the deep bound state is outside the domain of the EFT,
while the physics of the shallow bound state is correctly described
by the EFT.

The field $T$ can be used to describe the shallow bound state in processes
other than $\psi\psi$ scattering, if one keeps in mind
that the residue of $-iS$ at the pole is not 1,
but $((g^{(R)})^2/g^2)(\sigma/(1+\sigma m^2 g^2/8\pi\sqrt{mB_s}))$.
Despite the resemblance, this is not an EFT with
an ``elementary'' field for the bound state. The dibaryon field
shows structure at $Q\gaprox \aleph$, which arises from 
$\psi$ loops. 
At $Q\ll \aleph$, the EFT (with or without dibaryon) can be matched
onto a lower-energy EFT containing a ``heavy'' field for the bound state
which does not couple to $\psi\psi$ states.
This can be seen by expanding $S$ for $p^0= -B_s + \delta p^0$,
in which case
\begin{equation}
S(p^0, \vec{p}\,) \rightarrow 
   \frac{i Z}{\delta p^0- \frac{\vec{p}^{2}}{2(2m-B_s)} +i\epsilon}
\end{equation}
\noindent
with a wave function normalization
\begin{equation}
Z= -2\sigma \frac{(g^{(R)})^2}{g^2} \sqrt{\frac{B_s}{B_d}},
\end{equation}
\noindent
up to terms suppressed by $Q/\aleph$.

\subsection{Another unnatural EFT}

Without fine-tuning, the scattering length comes out of natural size. In
the previous subsection we saw how, adjusting the potential to produce
a shallow bound state, the scattering length results big.
Let us consider now the remaining case: an unnaturally 
small scattering length. This happens when a parameter $\alpha$ 
of the underlying theory is very close to the value $\alpha_c'$
that generates a zero of the amplitude right at threshold.
Here again, we are dealing with two scales: the natural scale $M$ and 
a small scale $\aleph' = |\alpha/\alpha_c'-1| M \ll M$;
it is convenient to introduce an alternative 
small scale $\Omega= \sqrt{|\alpha/\alpha_c'-1|}M \ll M$.
I will consider the case of an $S$-wave shallow zero
by taking
$C_{0}^{(R)}= 4\pi \gamma_{0} \aleph'/m M^2 =4\pi \gamma_{0} \Omega^2/m M^3$, 
and $C_{2n}^{(')(R)}= 4\pi \gamma_{2n}^{(')}/m M^{2n+1}$ for $n\geq 1$,
with 
$\gamma_{2n}^{(')}$ dimensionless
parameters of $O(1)$. 
This of course 
recovers the natural scenario when $\alpha$
is tuned out of $\alpha_c'$, and
 $\Omega$ becomes comparable to $M$.

In this case, the loop expansion in Eq. (\ref{Tpert}) is
in $m Q C_0^{(R)}/4\pi\sim Q\aleph'/M^2 =(Q/\Omega) (\Omega/M)^3$.
The derivative expansion
in other channels behaves as in the natural case,
but in the $S$-channel it is (again)
less trivial than in the natural scenario:
$C_{2n+1}^{(R)} Q^2/C_{2n}^{(R)} \sim Q^2/M^2$ 
for $n\geq 1$, yet
$C_2^{(R)} Q^2/C_0^{(R)}\sim Q^2/\aleph' M = (Q/\Omega)^2$.

For $Q\ll \Omega$ both  expansions are still perturbative:
the leading order is one $C_0$ contact interaction,
with corrections 
at $(Q/\Omega) (\Omega/M)^3$ from two $C_0$'s,
at $(Q/\Omega)^2$ from $C_2$ and $C_2'$, and so on.
This perturbative expansion is similar to the natural case.

\begin{figure}[t]
\centerline{\psfig{figure=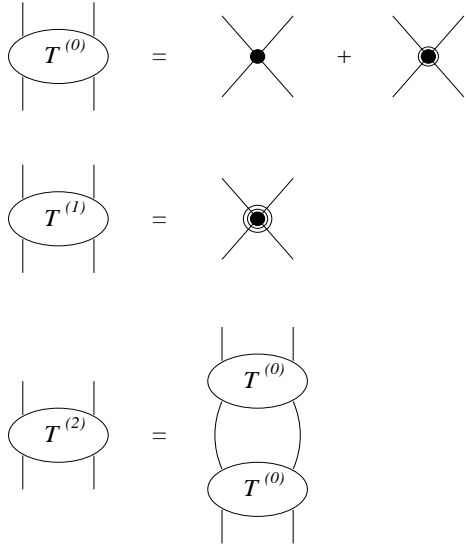,height=2.85in,width=2.40in}}
\caption{First three orders of the
two-particle amplitude $T$ in an EFT with a shallow zero.
Two solid lines represent a Schr\"odinger propagator;
a heavy dot stands for a $Q^0$ contact interaction, 
a dot within a circle for a $Q^2$ contact interaction, and
a dot within two circles for a $Q^4$ contact interaction.
\label{fig:unnat2t}}
\end{figure}

When $Q\sim \Omega$, however, the $C_2$ and $C_2'$ terms become comparable to 
the  $C_0$ term. The leading order, shown in 
Fig. \ref{fig:unnat2t}, is then simply 
\begin{equation}
T_{os}^{(0)}(k, \hat{p}\,'\cdot \hat{p})  = -\left( C_0 + 2 C_2 k^2 
                     +2 C_2' k^2  \hat{p}\,'\cdot \hat{p}\right), 
                                                        \label{T'zero}
\end{equation}
\noindent
which is of $O((4\pi\Omega^2/m M^3)(1+(Q/\Omega)^2))$.

Further derivative terms come in powers of $\aleph'/M =(\Omega/M)^2$,
so that the first corrections come from four-derivative contact terms:
$C_4$, and other $P$- and $D$-wave terms. Waves higher
than $S$ again behave as in the natural case, so they are not very
interesting.
I write
\begin{eqnarray}
T_{os}(k, \hat{p}\,'\cdot \hat{p}) & = & 
       T_{os}^{(0)}(k, \hat{p}\,'\cdot \hat{p}) 
       +\dots\nonumber \\
 & = & (T_{os}(k))_0 
       -2C_2^{'(R)} k^2 \hat{p}\,'\cdot \hat{p}
       +O((4\pi\Omega^2/m M^3)(\Omega/M)^2), \nonumber \\
\end{eqnarray}
\noindent
and focus on the $S$-wave component $(T_{os}(k))_0$.

The first $S$-wave correction is just
\begin{equation}
T_{os}^{(2)}(k)  =  -4C_4 k^4 
      \label{T'1},
\end{equation}
\noindent
which is of $O((Q/\Omega)^2 (\Omega/M)^2)$
relative to $T_{os}^{(0)}(k)$ in Eq. (\ref{T'zero}).
The second corrections come from the one-loop graphs with $C_0$ and 
$C_2$ vertices,
\begin{equation}
T_{os}^{(3)}(k)  =  
      - \left[ (C_0 + C_2 k^2)^2 I_0(k)  
     + 2 C_2 (C_0 + C_2 k^2) I_2(k) 
     + C_2^2 I_4(k) \right],
     \label{T'2}
\end{equation}
\noindent
which are of $O((Q/\Omega)(\Omega/M)^3(1+(Q/\Omega)^2+(Q/\Omega)^4)$
relative to $T_{os}^{(0)}(k)$ in Eq. (\ref{T'zero}).
And so on.
See Fig. \ref{fig:unnat2t}.

The $S$-wave amplitude is
\begin{eqnarray}
(T_{os}(k))_0 & = & 
       T_{os}^{(0)}(k)+  T_{os}^{(2)}(k) 
       +T_{os}^{(3)}(k) +\dots\nonumber \\
& = & 
  -\left[ C_0^{(R)} + 2 C_2^{(R)} k^2 +4C_4^{(R)} k^4
  -\frac{imk}{4\pi} (C_0^{(R)} + 2 C_2^{(R)} k^2)^2 +\ldots \right]\nonumber \\
 & = & - \left[ \frac{1}{C_0^{(R)}+ 2 C_2^{(R)} k^2} 
      \left(1- \frac{4C_4^{(R)} k^4}{C_0^{(R)}+ 2 C_2^{(R)} k^2}\right)
               +\frac{imk}{4\pi} \right] ^{-1}   \nonumber \\
 & & \ \ \ \ \left(1+O((\Omega/M)^4)\right)
                                                   \nonumber \\
 & = & - \left[ \frac{1}{C_0^{(R)}+ 2 C_2^{(R)} k^2 +4C_4^{(R)} k^4}
                +\frac{imk}{4\pi}\right] ^{-1} 
\left(1+O((\Omega/M)^4)\right).        \nonumber \\        
\label{renT'on}
\end{eqnarray}
\noindent
As before, cut-off dependence of integrals and parameters was lumped 
together in renormalized parameters. 
For simplicity in writing the higher-order terms, I here took 
$Q\sim \Omega$.

One can see that Eq. (\ref{renT'on}) 
is equivalent to a power expansion of $k^{-1} \tan \delta$, and
comparison with Eq. (\ref{invere}) shows that the effective range parameters
have the same expressions as in 
Eqs. (\ref{sscatlen})--(\ref{sshaparam}).
There is no change in the scaling of $l(>0)$-wave parameters with $M$,
but the $S$-wave is once again modified. 
The scattering length $a_0$ now scales with
$M^3/\Omega^2$ rather than $M$, $a_0= O(\Omega^2/M^3)$, 
the 
effective range $r_0$ scales with $\Omega^4/M^3$ rather than $M$, 
$r_0 =O(M^3/\Omega^4)$,
the shape parameter $P_0 r_0^3$ with $\Omega^6/M^3$ rather than $M^3$,
$P_0 =O(\Omega^6/M^6)$, and so on. 
That this is the correct scaling of the effective range parameters will
be shown in the case of the square-well potential in Sect. 6.

Alternatively, we can rewrite Eq. (\ref{renT'on}) as
\begin{equation}
(T_{os}(k))_0= \frac{4\pi/m}{-\frac{R}{1+ Pk^2} -A +\ldots -ik} 
\end{equation}
\noindent
where each of the parameters $R, P, A, \ldots$ has an expansion
in powers of $(\Omega/M)^2$:
\begin{eqnarray}
R &=& \frac{4\pi}{m C_0^{(R)}} \left(1+\frac{C_0^{(R)}C_4^{(R)}}{(C_2^{(R)})^2}
                                +\ldots \right), \nonumber\\
P &=& 2\frac{C_2^{(R)}}{C_0^{(R)}}
       \left(1-\frac{C_0^{(R)}C_4^{(R)}}{(C_2^{(R)})^2}
                                +\ldots \right), \nonumber\\
A &=& -\frac{4\pi}{m C_0^{(R)}} \frac{C_0^{(R)}C_4^{(R)}}{(C_2^{(R)})^2} 
                                (1+ \ldots ), \nonumber\\
& \ldots &               \label{RPA}
\end{eqnarray}
\noindent
They are $R\sim M^3/\Omega^2 \gg M$, $P\sim 1/\Omega^2 \gg 1/M^2$,
$A\sim M$, $\dots$
Clearly, the effect of this kind of fine-tuning is to generate
a shallow pole in $k \cot \delta$, at $k=O(\Omega)$.
This not only forces the amplitude to be small right at threshold 
---which translates into small $a_0$--- 
but also imparts huge energy dependence to the amplitude near
threshold 
---which translates into large $r_0$. 
The usual
effective range expansion is confined to very small momenta $Q \ll \Omega$. 
Eq. (\ref{RPA}) generalizes this expansion to
momenta which include
the position of the pole.

As $Q$ increases, $Q\gaprox \Omega$, the $C_2$ term becomes dominant,
and the most important corrections come from repeated insertions
of $C_2$, generating powers of $Q/M$. The EFT expansion
becomes progressively worse  until
it breaks down at $M$.

\subsection{Moral}

In summary, I have shown in the subsections above  
that, in the case of short-range interactions,
{\it the EFT approach, when applied consistently, is 
completely equivalent to the (generalized) effective range expansion}.
The details of how this works depend on the power counting,
which in turn depends on whether we are considering a
straightforward natural underlying theory,
an unnatural theory with a shallow pole,
or an unnatural theory with a shallow zero.
In any case,
if the EFT parameters are known from the underlying, more 
fundamental theory, then the effective range parameters can be predicted; 
otherwise, the EFT parameters can be determined by fitting $\psi\psi$ 
scattering data with the effective range expansion.  
We will consider a toy example in Sect. 6.
Before that, we will explore the connection with the 
Schr\"odinger equation.

\section{Pseudo-potential and boundary conditions}

The system we have been considering ---that of two heavy, stable
particles with
short-range interactions--- is traditionally dealt with in
the framework of non-relativistic quantum mechanics,
where a solution is attained by solving the Schr\"odinger 
equation with a potential believed to describe the short-range
dynamics ---a model.
If no approximations are used in the solution, 
this procedure is equivalent to obtaining the full 
scattering amplitude
by iteration of the potential to all orders. 
The potential is naturally defined 
as the sum of diagrams which do not contain $\psi$ poles.
For short-range forces, the bare 
potential coincides ---apart from a phase---
with the vertex (\ref{ver}). 
An expansion of the 
bare potential is
essentially an expansion of the Lagrangian itself. 
I have shown that the leading part of the two-particle
amplitude can be obtained from the leading term(s) 
of the bare potential (\ref{ver}) and
from the Schr\"odinger propagator (\ref{sprop}),
followed by renormalization. 
Renormalization is the price paid for 
the model-independence of the EFT approach. 

One might wonder what (if any) Schr\"odinger equation is generated by 
the renormalization procedure.
This is particularly relevant
because in Ref. \cite{cohen}  
it was argued ---following Wigner \cite{Wigner}--- 
that $r_0$ cannot be positive if we
take a sequence of ever shorter-range, (possibly non-local) hermitian
potentials in a coordinate space Schr\"odinger equation.
This procedure can be thought of as a particular regularization of
the Schr\"odinger equation with a delta-function potential,
followed by removal of the regulator.
This suggests that the Schr\"odinger equation for the 
``renormalized potential'' (if it exists) 
violates the assumptions made in Ref. \cite{cohen} 
about the nature of the potential. 

I am going to show now that the ``renormalized" Schr\"odinger 
equation ---{\it i.e.} the equation in terms of the observable,
renormalized parameters---
contains the pseudo-potential discovered long ago in 
Refs. \cite{Peierls,Breit,Huang}. This, in turn, is equivalent to a normal 
Schr\"odinger equation with unusual boundary conditions 
at the origin \cite{Fermi,Breit}.
 
I have shown in Sect. 3 that what is to be taken as
the leading part of the potential and whether this leading
piece has to be
iterated to all orders depend on fine-tuning (or lack thereof)
in the underlying theory. 
For simplicity in this section I will 
{\it (i)} leave implicit which terms
should be iterated to all orders, and which ones can be treated 
perturbatively;
and
{\it (ii)} neglect 
relativistic corrections, which we have seen are less important than
the  $Q^2$ four-$\psi$ interaction in both fine-tuning scenarios.
I will also concentrate on the $S$-wave, inclusion of other waves
following analogous arguments.

The bare Schr\"odinger equation can be obtained by acting with the operator
$E- H_0$ on the (scattering) 
wave-function $|\Psi\!>$ , which in coordinate space is,
asymptotically,
\begin{equation}
\Psi(\vec{r})= \exp(i\vec{p}\cdot \vec{r}) + T_{os}(k) G_0(r;k)  \label{Psi}
\end{equation}
\noindent
with $|\vec{p}\:|=k$ and $G_0(r;k)$ given by Eq. (\ref{spropcoord}).
Not surprisingly we find
\begin{equation}
<\!\vec{r}\:|(E- H_0)|\Psi\!>= <\vec{r}\:|v_{os}|\Psi\!>,   \label{bareSch}
\end{equation}
\noindent
where
\begin{equation}
v_{os}(k)= C_0+ 2 C_2 k^2 + \dots               \label{barev}
\end{equation}
\noindent
in momentum space.

As we saw, the rhs of Eq. (\ref{bareSch}) does not make sense as it stands.
The effects of renormalization on the on-shell amplitude are 
two-fold. One is to replace $v(p,p')$ by
\begin{equation}
v^{(R)}(k)= C_0^{(R)}+ 2 C_2^{(R)} k^2 + \dots             \label{vosR}
\end{equation}
\noindent
The other is to substitute the (free) Schr\"odinger propagator $G_0(p;k)$ 
by
\begin{equation}
G_0^{(R)}(p;k)= \frac{-m}{p^2 -k^2 -i\epsilon} D(k^2/p^2), 
\end{equation}
\noindent
where $D(k^2/p^2)$ is such that
\begin{equation}
\int \frac{d^3l}{(2 \pi)^3} \: G_0^{(R)}(l;k) =
-\frac{m}{2\pi^2}\int_0^{\infty}dl \:
          \frac{l^2 D(k^2/l^2)}{l^2 -k^2 -i\epsilon} =
-\frac{imk}{4\pi}.       \label{Dcon}
\end{equation}

Eq. (\ref{Dcon}) does not define $D(k^2/p^2)$ uniquely. The simplest solution
is
\begin{equation}
D(k^2/p^2)= \frac{k^2}{p^2},
\end{equation}
\noindent
and the corresponding coordinate-space propagator is
\begin{equation}
G_0^{(R)}(r;k)= G_0(r;k) - G_0(r;0).
\end{equation}
\noindent
Another solution 
is 
\begin{equation}
D(k^2/p^2)= \frac{2k^2}{p^2 -k^2 -i\epsilon},
\end{equation}
\noindent
giving
\begin{equation}
G_0^{(R)}(r;k)= \frac{\partial}{\partial r} (r G_0(r;k)).
\end{equation}
An infinite number of other solutions exist, but their important common
feature is that they soften the $r\rightarrow 0$ behavior of the propagator:
as it is obvious from the defining Eq. (\ref{Dcon}),
\begin{equation}
G_0^{(R)}(r;k)=  -\frac{imk}{4\pi} (1+ O(kr)).            \label{GR}
\end{equation}

Renormalization therefore defines the rhs of Eq. (\ref{bareSch}) as
\begin{equation}
<\!\vec{r}\:|v_{os}|\Psi\!> \equiv <\!\vec{r}\:|v^{(R)}|\Psi_R\!>,   
\label{vvR}
\end{equation}
\noindent
where
\begin{equation}
\Psi_R(\vec{r}) = \exp(i\vec{p}\cdot \vec{r}) + T_{os}(k)  G_0^{(R)}(r;k) 
\end{equation}
satisfies
\begin{equation}
\delta(\vec{r})\Psi_R(\vec{r}) =
  \delta(\vec{r})\frac{\partial}{\partial r} (r \Psi(\vec{r})). \label{PsiR}
\end{equation}
\noindent
due to Eq. (\ref{GR}).
We can now define an ``effective renormalized potential'' $\hat{v}^{(R)}(k)$ 
acting on the original wave-function,
\begin{equation}
<\!\vec{r}\:|v^{(R)}|\Psi_R\!> \equiv <\!\vec{r}\:|\hat{v}^{(R)}|\Psi\!>,  
\label{vRveff}
\end{equation}
\noindent
finding
\begin{equation}
\hat{v}^{(R)}(\vec{r}) = (C_0^{(R)}+ 2 C_2^{(R)} k^2 +\dots) \delta(\vec{r})
                       \frac{\partial}{\partial r} r.
\end{equation}
\noindent
This is the ``potential'' which encodes all the effects of renormalization:
used with free
Schr\"odinger propagation it produces 
by construction the correct, renormalized amplitude.
That is, the renormalized Schr\"odinger equation is
\begin{equation}
-\frac{1}{m} (\nabla^2 +k^2) \Psi(\vec{r})= 
 -(C_0^{(R)}+ 2 C_2^{(R)} k^2 +\ldots) \delta(\vec{r})
 \frac{\partial}{\partial r} (r \Psi(\vec{r})).   \label{renSch}
\end{equation}
It is easy to invert the preceding reasoning and check that this 
Schr\"odinger equation
indeed leads to the on-shell $T$-matrix obtained previously.

\begin{figure}[t]
\centerline{\psfig{figure=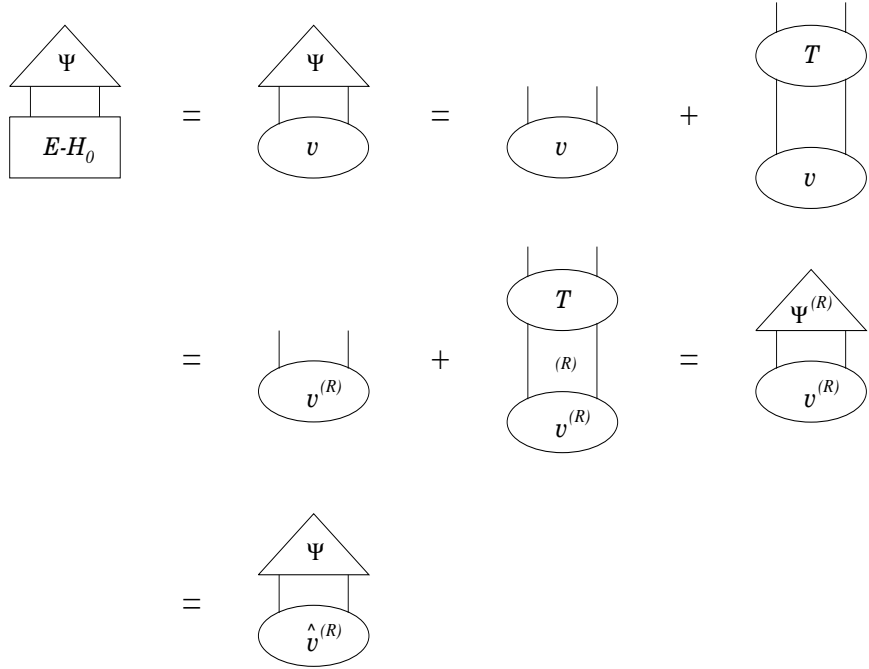,height=3.5in,width=4.50in}}
\caption{Diagrammatic derivation of the pseudo-potential:
the first line is ill-defined; the second line follows from
renormalization; the third line is the definition of 
the pseudo-potential $\hat{v}^{(R)}$.
\label{fig:sch}}
\end{figure}

The above derivation can be sketched in diagrammatic terms ---see 
Fig. \ref{fig:sch}.
The steps leading to 
Eq. (\ref{renSch}) 
can be repeated starting with the bound state wave-function
\begin{equation}
\Psi(\vec{r})= -\frac{N}{m} G_0(r;i\kappa).  \label{Psibound}
\end{equation}
\noindent
($N$ is a normalization factor.)

Recalling the connection (\ref{sscatlen}), (\ref{seffran}), {\it etc.},
between EFT and effective range 
parameters, we see that this is nothing but 
a generalization of the well-known pseudo-potential of Fermi and Breit 
\cite{Fermi,Breit,Huang}. Therefore,
{\it renormalization of an EFT with only short-range interactions leads 
to a Schr\"odinger equation with a pseudo-potential determined by 
renormalized parameters}, rather than 
a Schr\"odinger equation with a local, 
smeared delta-function-type potential determined by bare parameters.

Further insight into the effect of renormalization can be gained
by looking at
the solution of Eq. (\ref{renSch}). 
Recall that an ordinary potential does not generate 
an $1/r$
behavior of the wave-function close to the origin because action of the 
Laplacian on such singularity produces a delta function:
$\nabla^2 (1/r) = - 4\pi \delta(\vec{r})$. A local delta-function
potential ---as our bare potential (\ref{barev})--- 
in turn is too singular to admit a solution of this type because then
the potential side of the Schr\"odinger equation becomes
$v \Psi \sim \delta(\vec{r})/r$, more singular than
the kinetic side. (This is the quantum-mechanical manifestation
of 
our original divergence problem.) The effect of the 
$\frac{\partial}{\partial r} r$ operator in the pseudo-potential is to
soften this singularity at the origin,  
replacing $\delta(\vec{r})/r$ by $\delta(\vec{r})$. 
Indeed, we can find the solution of Eq. (\ref{renSch}) by expanding
$\Psi(\vec{r}) = \sum_{n=-\infty}^{+\infty} A_n r^n$ at small $r$.
The result is completely determined by $A_{-1}$ as
\begin{equation}
\Psi(\vec{r}) = A_{-1}\left[ \frac{1}{r}- 
             \frac{4\pi}{mC_0^{(R)}} 
             \left(1- 2 \frac{C_2^{(R)}}{(C_0^{(R)})^2} k^2 +\ldots\right) 
             + O(r)\right].
\end{equation}

Now, the same solution follows from the free Schr\"odinger equation,
\begin{equation}
-\frac{1}{m} (\nabla^2 +k^2) \Psi(\vec{r})= 0,
\end{equation}
\noindent
with a peculiar boundary condition at the origin,
\begin{equation}
\left. \frac{\partial}{\partial r} \ln (r \Psi(\vec{r}))\right|_{r=0} =
          -\frac{4\pi}{mC_0^{(R)}} 
      \left(1- 2 \frac{C_2^{(R)}}{(C_0^{(R)})^2} k^2 +\ldots\right). \label{bc}
\end{equation}
\noindent
Again, using Eqs. (\ref{sscatlen}), (\ref{seffran}), {\it etc.},
we recognize the first term in Eq. (\ref{bc}) as the boundary 
condition of Breit \cite{Breit}. {\it The renormalized EFT of only 
short-range interactions is equivalent to an energy expansion of the most 
general condition on the logarithmic derivative of the wave-function at the 
origin.}

The physics behind these Schr\"odinger formulations is clear if
we look at the resulting radial wave-functions $r\psi(r)=u(r)$.
{\it (i)} For $k^2=-\kappa^2<0$,
\begin{equation}
u(r)=c_{(-)} e^{-\kappa r},
\end{equation}
\noindent
with $c_{(-)}$ a normalization factor and $\kappa$ given by 
Eq. (\ref{boundmom}).
{\it (ii)} For $k^2\geq 0$,
\begin{equation}
u(r)=c_{(+)}\left(\sin kr - \frac{a_0k}{1-\frac{a_0r_0 k^2}{2}} \cos kr\right),
\end{equation}
\noindent
with $c_{(+)}$ a normalization factor and $a_0$ and $r_0$ given in 
Eqs. (\ref{sscatlen}) and (\ref{seffran}).
By changing the unspecified dependence of $c_{(+)}$ on $k$ and $\kappa_{s,d}$
we do not change the underlying pole structure of the $\psi\psi$ amplitude,
but we do change the Jost functions. 
This is a reflection of the fact that the EFT cannot distinguish
among short-range potentials which produce the same asymptotic behavior. 
Indeed, it is not difficult to see that with appropriate choices  of $c_{(+)}$
we can cover all situations considered in Ref. \cite{Bargmann}.
The EFT shrinks to the origin the complicated wiggles that the ``underlying''
wave-function might have at small distances, thus replacing
them with a smooth behavior that matches onto the tail of the  underlying
wave-function. The effective wave-function is all tail for the tail
is all one can see from far away.

\section{Potential, Regularization issues}

I have so far avoided using specific regularization schemes in order
to emphasize the features of the renormalized theory, which is all that
determines observables anyway. 
The power countings derived in Sect. 3 are valid independent
of the details of the regularization scheme adopted.
The conclusion that the EFT approach is completely equivalent to the 
effective range expansion, the method of pseudo-potentials, and to analytic 
boundary conditions had previously been hidden by the use of specific
regulators. 
The issue of regularization has also clouded the relation to
the traditional approach of using a potential in the Schr\"odinger
equation. 
Here a few words are spent on these related issues: 
the roles of a potential and of regularizations schemes.
(Some of these remarks have been presented in Ref. \cite{vKolckn},
and been shown since to be supported by explicit numerical
investigation \cite{steele}.)

In the case of a natural theory, it is clear that iteration of a 
potential is superfluous. Within its region of validity, the EFT
is perturbative: nothing needs to be iterated to all orders,
and any bound states, if they exist, exist beyond the range of the
EFT. 
Iteration of the potential includes only part of the higher-order
terms, and this incomplete set cannot {\it a priori} be taken seriously.  
Indeed, it has been shown that the iteration
of momentum-independent contact interactions
does not reproduce
the non-analytical behavior of a toy natural underlying theory
\cite{luke2}.
On the other hand, since the error induced by iteration
is small, it does not affect observables much and does no harm
---as long as one remembers that higher orders have not been accounted for
correctly. 

The more interesting cases are the ones with fine-tuning
in the underlying theory. For definiteness I focus on the case
of a shallow bound state, which has attracted a lot of attention lately
in the context of nuclear forces. 

The appropriate formula for the amplitude, 
correct up to and including $O((\aleph/M)^2)$,
is Eq. (\ref{renTon3}), which I obtained
using perturbation theory for the corrections to the $Q^0$ interaction in the
bare potential (\ref{ver}).
The same result can 
be obtained more easily by first summing 
the bubbles with the full $v(p',p)$ and then expanding in powers of the 
energy. For example, keeping track only of the $Q^2$ terms 
in the $S$-wave,
\begin{equation}
(T_{os}(k))_0 = -\left(\frac{1}{C_0+2C_2 k^2}- I(k)\right)^{-1},  \label{Ton}
\end{equation}
\noindent
with
\begin{equation}
I(k) = \left(1+\frac{k^2}{2m^2}\right) I_0(k) 
           - \left(2\frac{C_2}{C_0}+\frac{1}{4m^2}\right)
             \frac{m L_3}{2\pi^2}.
\end{equation}
\noindent
This corresponds to the standard procedure of iterating the full potential,
but
it is now obvious that to remove the infinities we {\it have to} expand
\begin{equation}
\frac{1}{C_0+2C_2 k^2}= \frac{1}{C_0}-2 \frac{C_2}{C_0^2} k^2+\ldots, 
\label{exp}
\end{equation}
\noindent
and consistently neglect $Q^4$ terms. 
We are then back to Eq. (\ref{renTon3}).
Terms of higher order in the $Q/M$ expansion can only be
made regulator-independent by renormalizing higher order 
parameters in the Lagrangian. Indeed, as we have seen before,
$O(Q^4)$ terms in $v(p,p')$
will contribute $k^4$ terms to Eq. (\ref{Ton}).

Regulator-dependence is allowed insofar as it is small, that is,
of the same magnitude as smaller terms that have been neglected.
In practice, in the perturbative treatment of corrections
some cut-off dependence 
will remain from $R(k^2/\Lambda^2)$, for it contains
arbitrarily high powers of $Q/\Lambda$. 
Now, what we have just done in Eq. (\ref{Ton})
was to re-sum some sub-leading terms, then to
throw them away in Eq. (\ref{exp}). 
The first step is carried out by iterating the potential,
but the last step requires extra work. Omitting it
introduces further cut-off dependence that creeps in through
bare parameters rather than $R$; the size
of the error induced by such laziness depends on the 
relative size of the bare parameters.

For example, to $O((\aleph/M)^2)$, 
by solving the the Schr\"odinger equation with a finite cut-off
we induce a shape parameter
\begin{equation}
P_0 r_0^3 = \frac{16\pi C_2^2}{mC_0^3}
            +\frac{2}{\pi\Lambda^3}
             \left( R'(0) +\frac{\Lambda^2}{2m^2}R(0)\right)
            +\frac{r_0}{8m^2}.
\label{fakeshape}
\end{equation}
\noindent
We see that apart from a regularization-independent relativistic correction,
cut-off dependence enters through both 
$R$ and the bare parameter ratio $C_2^2/C_0^3$.

The induced error in the amplitude
coming from $R$ is $O(\aleph/\Lambda)$ and will be 
small as long as 
$\Lambda \gaprox M$.
It can in particular be completely removed 
by taking $\Lambda\rightarrow \infty$.
It is natural to ask whether there are regulator schemes in which
we can forfeit the last step in Eq. (\ref{exp}) 
and still manage to remove the errors coming from bare parameters, too. 
In more complicated situations dealt with in the literature
---where there 
exists a long-range potential which can only be solved numerically---
both of these spurious effects can be present and
impossible to remove explicitly. It is desirable
to find schemes that at least minimize these 
spurious effects; yet, we must be prepared
to discover that not all schemes are flexible enough to achieve this.

To any given order, the equations 
relating bare and renormalized parameters will be truncated,
and beyond leading order, be highly non-linear ---see
Eqs. (\ref{1/C0R}), (\ref{C2R}), for example.
Inverting them, we can find the bare parameters in terms of
the renormalized parameters (known from data) and the cut-off:
$C_{2n}=C_{2n}(C_{2n}^{(R)}; \Lambda)$.
This of course means that the bare parameters are not observables,
depending on the scheme chosen. 
Given $C_{2n}^{(R)}$'s,
whether one finds real solutions $C_{2n}$ for any $\Lambda$ depends on the 
actual values of the $\theta_{n}$'s in Eq. (\ref{divs}).
In particular, 
to $O((\aleph/M)^2)$ the important term for large $\Lambda$
is $C_2^{(R)}\theta_3\Lambda^3$; 
the sign of $C_2^{(R)}$ being that of
$r_0$, it is the sign of $\theta_3$ which is of concern.
It was pointed out in Ref. \cite{cohen} 
that the most obvious cut-off schemes
(such as a sharp momentum-space cut-off and a coordinate-space
regularization by square wells) imply $\theta_3 >0$.
Imposing real bare parameters, it is found that
$r_0 \leq 2/\Lambda$ for large $\Lambda$, or in other words, that the limit
$\Lambda \rightarrow \infty$ can only be taken if $r_0\leq 0$.
This is a reformulation in  field-theoretical language of Wigner's theorem
\cite{Wigner}.
  
While such a constraint is unusual, it is doubtful that it is of much 
relevance to EFTs applied consistently to a certain order. 
First, this is obviously a regularization-scheme-dependent issue.
Dimensional regularization with minimal subtraction
has the feature that $\theta_{n}=0$,
and it is easy to construct (smooth) cut-off schemes such that 
$\theta_3 \leq 0$ \cite{david3}. 
In the worst case scenario, 
the above constraint is a result of attempting to remove the
cut-off in an inadequate 
regularization scheme. 
But actually this constraint only arises from 
a constraint imposed on unobservable bare parameters. No matter
what process one is considering in the low-momentum regime, 
the bare parameters never appear, except in the right 
combinations with the cut-off to produce renormalized parameters:
an imaginary bare parameter is of no relevance.
Finally, even if one insists on the unfortunate choice of regulator 
and the reality constraint on bare parameters, a positive $r_0$ can still 
be achieved by keeping the cut-off finite and of the order of the 
mass scale of the underlying theory, $\Lambda\sim M$. 

Regardless, keeping a cut-off $\Lambda\sim M$ has one useful consequence. 
In this case, the large ---containing inverse powers of $\aleph$---
renormalized parameters $C_{2n}^{(R)}$ can be generated by
natural-size bare coefficients
$C_{2n}=O(4\pi /m M^{2n+1})$.
For example, 
$C_0=-(2\pi^2/ \theta_1 m \Lambda) (1+ \pi \aleph/2\gamma_0 \theta_1 \Lambda
+\ldots)$ is of natural size
even though
$C_0^{(R)}=4\pi\gamma_0/m \aleph$ is large.
Similarly,
$C_2=(\pi \gamma_2/2 \theta_1^2 \gamma_0 m \Lambda^2 M) 
\linebreak
(1+ \pi \aleph/\gamma_0 \theta_1 \Lambda +\ldots)$. 
What a cut-off does is to give a scale for the bare parameters to be 
fine-tuned against.
This way the ratio of bare parameters can be made of natural
size as well. The difference between a truncation of the rhs of
Eq. (\ref{exp}) and the lhs is then no bigger than other neglected
small effects such as higher-order interactions in the Lagrangian: 
for example, the shape parameter (\ref{fakeshape}) induced
by the cut-off is in this case $O(1/M^3)$.
{\it A power expansion of the bare potential followed
by its iteration to all orders is under these circumstances no worse
than a perturbation treatment of the corrections
to leading order.} 

We actually can invert the previous reasoning and use the freedom
introduced by the cut-off to save some work. 
Given a regularization procedure, we can further fine-tune $\Lambda$ 
to its ``optimal" value that improves agreement with data.
For example, if we are working to  $O((\aleph/M)^2)$,
we can fine-tune $\Lambda$ so 
that Eq. (\ref{fakeshape}) fits the experimental value
for $P_0$; such a procedure was considered in Ref. \cite{cohen} in the case
of the $^1S_0$ $NN$ phase shift. 
If we are dealing with a single channel, 
this is a consistent procedure: it simply means 
that $O(Q^4)$ contributions have been included, but they do not need
to be written explicitly because the corresponding bare parameter is zero.
In general, however, only part of the higher order effects can be accounted 
for in this manner.

The above discussion holds for any scheme where a cut-off mass scale 
appears explicitly. 
Dimensional regularization with
minimal subtraction,
first applied to this problem
in Ref. \cite{Kaplan2}, is special because it defines
the integral (\ref{regbubbleany}) with $L_n=0$ and $R=0$.
It suffers from no errors generated by $R$, but
it forces the bare parameters to coincide with
the renormalized ones and thus be unnaturally large.
Because no cut-off dependence is visible,
if one works to $Q^2$ terms it is not immediately apparent that  
$1/(C_0 + 2C_2 k^2)$ has necessarily to be expanded in $k^2$. 
It was one of 
the results of Ref. \cite{Kaplan2} that a fitting of the $^1S_0$ $NN$ phase 
shift based on the effective range expansion worked much better than
one based on the form $1/(C_0 + 2C_2 k^2)$, and a
suggestion
was made that the most useful expansion is that of $T^{-1}$ rather than $v$.
One can understand the failure of the $1/(C_0 + 2C_2 k^2)$ fit by noticing that
the induced spurious shape parameter (\ref{fakeshape}) 
is here $P_0 r_0^3= a_0 r_0^2/4 -3r_0/8m^2 +1/4a_0m^4$,
which is $O(1/\aleph M^2)$. 
Since there is no regulator to blame the fine-tuning on, 
the fine-tuning contaminates
all other induced terms as well;
they are all automatically large if $a_0$ is large.
In dimensional regularization with
minimal subtraction one {\it cannot} substitute the perturbative treatment
of corrections by the iteration of the whole potential without
reducing the range of applicability of the theory 
from $M$ to the ratio of parameters in the power expansion
of the contact interactions, which is $\sqrt{\aleph M}$.

The perturbative treatment of corrections can still be carried out
with dimensional regularization. This is true of 
minimal subtraction or other subtraction schemes, for example
the convenient one invented in Ref. \cite{Kaplan4}.
(The power counting of Sect. 3.2 \cite{vKolckn} was re-discovered
independently using this new subtraction in Ref. \cite{Kaplan4}.)
As originally found in Ref. \cite{david1}, the contamination problem 
of minimal subtraction can be 
avoided by using the dibaryon field,
which
in view of our regularization-independent arguments of
Sect. 3 is not surprising.

\section{Toy}

The point of using an EFT is to describe low-energy physics with
minimal assumptions about higher-energy behavior.
The results of this paper hold regardless of the details of the
physics that generates the short-range interactions. They will
arise from exchange of (possibly many) particles of sufficiently high mass,
perhaps in a theory without a small coupling constant.
It might happen that this physics can be equally well described by 
some potential of range
$R$ much smaller than the wavelength of the interacting
asymptotic particles.
Therefore we expect that the EFT results should be valid
in the particular case of 
simple quantum-mechanical potentials. 

Here I will illustrate some of the results of Sect. 3 about
the scaling of effective range parameters
with a simple attractive square-well potential of range 
$R$ and depth $V_0$
\begin{equation}
V(r)= -V_0 \theta(R-r).
\end{equation}

Defining a dimensionless variable 
\begin{equation}
\alpha \equiv \sqrt{mV_0} R 
\end{equation}
\noindent
it is easy to find the $S$-wave amplitude 
\begin{equation}
(T(k))_0=-i\left[e^{-2ikR} 
     \frac{\sqrt{\alpha^2+(kR)^2} \cot\sqrt{\alpha^2+(kR)^2} +i kR}
          {\sqrt{\alpha^2+(kR)^2} \cot\sqrt{\alpha^2+(kR)^2} -i kR}
                 -1 \right].
\label{squareT}
\end{equation}
\noindent
It then follows that 
\begin{equation}
a_0= R \left(1- \frac{\tan\alpha}{\alpha} \right) 
\end{equation}
\noindent
and 
\begin{equation}
r_0= R \left(1- \frac{R}{a_0\alpha^2} -\frac{R^2}{3a_0^2}\right);
\end{equation}
\noindent
expressions for the shape and other parameters can also be worked out but
are more complicated. 
The amplitude Eq. (\ref{squareT}) has
an interesting pole structure as a function of $\alpha$ \cite{Nussenzveig}.
In particular, it can be shown explicitly that all poles with $|k|<1/R$
are indeed on the imaginary axis.
These poles cross $k=0$ when $\alpha$ 
takes one of the values $\alpha_c= (2n+1)\pi/2$.
The zeros of Eq. (\ref{squareT}), on the other hand, cross $k=0$ when $\alpha$ 
takes one of the values $\alpha_c'= \tan\alpha_c'$.

For generic values of $\alpha =O(1)$, 
there is only one mass scale $M = 1/R$ in the problem:
$\tan\alpha/\alpha \sim 1$ and thus
$a_0 =O(1/M)$, $r_0 =O(1/M)$, and $\kappa =O(M)$. 
As expected, in this generic case the effective range parameters,
bound state momenta, and positions of the zeros are all given
by the natural mass scale $M$. 

As $\alpha-\alpha_c$ becomes small,
\begin{equation}
a_0= \frac{R}{\alpha_c (\alpha-\alpha_c)} 
      \left[1-\frac{(1-\alpha_c^2)}{\alpha_c} (\alpha-\alpha_c)
              + O((\alpha-\alpha_c)^2) \right],
\end{equation}
\begin{equation}
r_0= R \left[1-\frac{1}{\alpha_c} (\alpha-\alpha_c)
              + O((\alpha-\alpha_c)^2) \right],
\end{equation}
\noindent
and the pole is at $k=i\kappa$,
\begin{equation}
\kappa=\frac{\alpha_c (\alpha-\alpha_c)}{R}
       \left[1+\frac{(1-\alpha_c^2/2)}{\alpha_c} (\alpha-\alpha_c)
              + O((\alpha-\alpha_c)^2) \right].
\end{equation}
\noindent
If we identify $\aleph= |\alpha/\alpha_c-1|/R$, we see that indeed
the scaling of these parameters is as anticipated in Sect. 3, namely
$a_0= O(1/\aleph)$, $ r_0= O(1/M)$, and $\kappa=O(\aleph)$.

Likewise, as $\alpha-\alpha_c'$ becomes small,
\begin{equation}
a_0= -R \alpha_c' (\alpha-\alpha_c') 
      \left[1+\frac{(1+\alpha_c'^2+\alpha_c'^4)}{\alpha_c'^3} 
                (\alpha-\alpha_c')
              + O((\alpha-\alpha_c')^2) \right]
\end{equation}
\noindent
and
\begin{equation}
r_0= -\frac{R}{3 \alpha_c'^2 (\alpha-\alpha_c')^2}
      \left[1-\left(3+\frac{2(1+\alpha_c'^2+\alpha_c'^4)}{\alpha_c'^3}\right) 
            (\alpha-\alpha_c')
              + O((\alpha-\alpha_c')^2) \right].
\end{equation}
\noindent
Again, identifying $\Omega= \sqrt{|\alpha/\alpha_c'-1|}/R$, 
we see that indeed
$a_0=O(\Omega^2/M^3)$ and $r_0= O(M^3/\Omega^4)$,
as advertised in Sect. 3.

\section{Discussion and Conclusion}

I have
developed here power countings for EFTs with natural and unnatural
short-range interactions
that are manifestly independent of the 
regularization scheme.  
I have shown that the modern EFT method applied to the problem of 
short-range forces is, {\it after 
renormalization}, completely equivalent to the ancient effective range 
expansion, and to the methods of pseudo-potentials and boundary conditions.
Some of the possible pitfalls in the implementation of EFT method were also 
discussed. Why, then, bother with the EFT method?

EFTs are nowadays 
thought to be the rationale for the successes of field-theory based particle 
physics. We have therefore gained 
insight into the ancient techniques by seeing how they arise from 
renormalization in the modern conceptual framework without extra dynamical 
assumptions. But it can be argued that this conceptual gain is compensated 
by a practical loss, in that one can 
more simply stick to the ancient techniques
with the same observable results.

The gain is indeed marginal for the problem at hand, but there are
advantages to a field-theoretical framework when other processes in
the same energy scale are considered, and we want to
treat them all consistently, free of off-shell ambiguities. 
Power counting applies to these processes as well,
although it has to be enlarged to accommodate new operators associated
with other particles.
For example, recently the three-body system has been attacked 
using the power counting of Sect. 3.2. 
In the case of nucleon-deuteron scattering, successful
model-independent predictions can be achieved very easily
\cite{Bedaque}. 

Yet the gain is potentially 
much more significant in more general cases where there are other, lighter 
degrees of freedom that generate longer-range forces. One should 
recall the reason why particle physics found its EFT paradigm: 
symmetries play a dominant role in physics because they
restrict the form of S-matrix elements (and consequently also 
determine what the relevant low-energy degrees of freedom are), 
and a Lagrangian framework is by far the easiest way
to incorporate symmetries. The problem considered in this paper had very 
weak symmetry constraints ---essentially only on 
particle number and invariance under small Lorentz boosts. In more 
interesting problems, the role of symmetry is bound to be more important,
and the EFT method more convenient than the ancient techniques.
We only need to return to the $NN$ system to offer an example: at energies 
comparable to the pion mass the pion has to be retained as a degree of 
freedom, and (approximate) chiral 
symmetry is the most important constraint
on pion interactions.
At sufficiently low energies the inclusion of the pion is not mandatory,
but can be carried out with minor adaptations to the power counting
presented above \cite{Kaplan4}.
The EFT method of chiral Lagrangians is
the only known, {\it systematic} way to treat this spontaneously broken
symmetry, and it has been successful in dealing with low-energy 
phenomenology from pion-pion interactions to few-nucleon forces 
\cite{Bernard,vKolckn}.

\vspace{1cm}
\noindent
{\large\bf Acknowledgments}

\noindent
I thank R. Jaffe and E. Lomon for prodding the investigation of 
the connection between EFT and boundary conditions. 
Extensive discussions with P. Bedaque were indispensable for my understanding 
of $\aleph$ counting and uses of the trans..., er, dibaryon. 
Helpful conversations
with S. Beane, T. Cohen, R. Furnstahl, D. Kaplan, G. Rupak, and B. Serot,
are gratefully acknowledged.
This research was supported in part by NSF grant PHY 94-20470
and DOE grant DE-FG-03-97ER41014.

\pagebreak

\end{document}